\documentstyle{mn}
\begin{document}
\title[Magnetic torques]{Magnetic torques between accretion discs and stars}
\author[U. Torkelsson]{Ulf Torkelsson\\
Institute of Astronomy, Madingley Road, Cambridge CB3 0HA, United Kingdom}
\maketitle

\begin{abstract}
I show in this paper that two types of magnetic torques can appear in the 
interaction between an accretion disc and a magnetic accretor.  There is the
well-known torque resulting from the difference in angular velocity between
the accretion disc and the star,
but in addition there is a torque coming from the interaction
between the stellar magnetic field and the disc's own magnetic field.
The latter form of magnetic torque decreases in strength
more slowly with increasing radius, and will therefore dominate at large
radii.  The 
direction of the disc field is not determined by the difference in
angular velocity between the star and the disc as in the Ghosh \& Lamb model,
but is rather a free parameter.
The magnetic torque may therefore either spin up or spin down the star, and the
torque changes sign if the magnetic field in the disc reverses.  I suggest 
that this mechanism can explain the torque reversals that have been observed in
some disc-fed X-ray pulsars.
\end{abstract}

\section{Introduction}

The interaction between a magnetic star and a surrounding accretion disc is 
one of the least understood aspects of accretion.
At the same time it is of importance for understanding 
the spin evolution of objects as diverse as T Tauri-stars 
and X-ray pulsars.
The generally adopted model \cite{ghosh:lamb2} has been challenged by recent 
observations of X-ray pulsars (e.g. Nelson et al. 1997).  I will discuss how
the Ghosh \& Lamb model fits in with current thinking regarding the
angular momentum transport in accretion discs and show that a modification of
the Ghosh \& Lamb model is necessary.

T Tauri-stars come in two forms. Classical T Tauri-stars that show clear signs
of having accretion discs and are slow rotators, and weak-lined (naked)
T Tauri-stars that are fast rotators and lack the signature of an accretion
disc.  The weak-lined T Tauri-stars are believed to represent a later
evolutionary stage, when the star has got rid of its accretion disc.  Models of
the rotational evolution of the T Tauri-stars based on the magnetic interaction
between the disc and the star have been constructed by several groups (e.g
Armitage \& Clarke 1996, Cameron \& Campbell 1993, Cameron, Campbell \&
Quaintrell 1995).

At the other end of the range of magnetic accretors are the accreting neutron 
stars.  BATSE/CGRO has made it possible to obtain essentially daily  
measurements of the spin frequencies of
a number of the brightest X-ray pulsars.  Some of these systems are known
to possess accretion discs, although other objects in the group are
accreting directly from a stellar wind.  I will not pay any attention to the
latter 
objects in this paper.  
Nelson et al. \shortcite{nelson97} have found that
the X-ray pulsars are oscillating between periods of spin-up and spin-down.
The frequency derivatives are comparable 
(apart from the difference in the sign)
in the two states, and a typical system spends about the same amount of time
in each of the two states.  The length of time a system spends in one of the
states before switching to the other state differs widely between the different
systems, in the case of Cen X-3 it is 10 - 100 days, but 4U 1626-67 has
only been observed to switch once, and thus may stay for 10 years or more
in one state.

The standard model for the spin changes was presented by Ghosh \& Lamb 
\shortcite{ghosh:lamb3}.  
They describe the interaction of a dipolar stellar
magnetic field with a diffusive disc.  The dipolar magnetic field is 
penetrating the disc, which winds up the magnetic field in the toroidal
direction, because of the angular velocity
difference between the accretion disc and the star.
The co-rotation radius, where the 
angular velocities of the disc and the star are
the same,
plays a key role in this model.
The magnetic field lines penetrating the disc inside the co-rotation radius
spin up the star, whilst those penetrating the accretion disc outside
the co-rotation radius brake the star.  The spin evolution of the star is 
therefore the result of a balance between the angular momentum carried by the 
accreting matter from the disc to the star, the magnetic spin-up torque from
the accretion disc inside the co-rotation radius, and the magnetic
spin-down torque
from the accretion disc outside the co-rotation radius.  The position of the
inner edge of the accretion disc varies with the accretion rate such that it
moves closer to the star when the accretion rate increases.  Thus we
expect the star to spin up, or at least spin down more slowly, when the 
accretion rate, or equivalently the luminosity, is high.  

A complicated, and not yet fully solved problem is the position and physical
properties of the boundary layer between the accretion disc and the
stellar magnetosphere.  Ghosh \& Lamb 
\shortcite{ghosh:lamb2} presented solutions for the boundary layer, but instead
of solving for the magnetic field in the disc given a certain magnetic 
diffusivity they assumed a magnetic field and solved for the diffusivity.
Heptinstall \& Campbell \shortcite{campbell:heptinstall} and Brandenburg \& 
Campbell \shortcite{brandenburg:campbell} have recently presented solutions
for the magnetic field and velocity in the disc given a certain model for the
magnetic diffusivity.  The disc is terminated as the magnetic force becomes
stronger than the viscous force in the disc (cf. Campbell 1992, 1997).  The 
temperature and disc thickness increases dramatically close to the inner disc
edge, and the disc is subject of a viscous instability similar to the
Lightman-Eardley instability (Lightman \& Eardley 1974, Lightman 1974).

Ghosh \& Lamb \shortcite{ghosh:lamb3} calculated the torque acting on the
star by integrating over the surface of the disc.  This is without a doubt 
the most
convenient approach, but it does not provide much information on how the
torque is transferred to the star.  To understand this problem it is 
necessary to obtain a solution for the structure of the magnetosphere.  Simple
solutions assuming no flows through the magnetosphere have been presented by
Bardou \& Heyvaerts \shortcite{bardou:heyvaerts}.  
The neglect of a flow through 
the magnetosphere may be unrealistic though, and
Li, Wickramasinghe \& R\"udiger \shortcite{li96} 
have presented a class of funnel flow solutions in
which the angular momentum is 
carried by the matter.
Their solutions include a
slow magnetosonic shock, in case the flow is sub-Alfv\'enic, 
close to the stellar
surface.  The shock is forced to be close to co-rotation with the star, which
minimises the torque between the shock and the star.  In addition
the toroidal magnetic field at the shock is small, which keeps the torque 
between the shocked and
unshocked matter small.  The authors thus conclude that most of the angular 
momentum of the matter in the funnel has to be propagated back to the 
accretion disc.
Wang \shortcite{wang} presents two major objections to this result.  He points
out
that a small toroidal field is enough to produce a significant torque on the
star.  Furthermore if the angular momentum is propagated back to the disc,
it must be transported outwards through the disc by the usual viscous torque,
but he argues that the viscous torque is too weak to do this.


Safier 
\shortcite{safier} goes a bit further and puts into question whether 
the magnetic
field of a T Tauri-star
is a closed dipole field.  He notes that the magnetic activity 
of a T Tauri star will heat up the atmosphere of the star and thus
produce a corona and a stellar wind.  This stellar wind should not be
confused with the jets, that are typically much more massive, but it dominates
the magnetic field of the star, which thus
assumes an open topology as it is drawn out by the wind.  
The
open field lines are mainly radial so fewer field lines penetrate the 
accretion disc than for a dipole magnetic field.

The large uncertainties in the models make it interesting to try to
simulate the interaction between the accretion disc and the
magnetosphere numerically.  Such simulations are in themselves sensitive to
the assumptions used to construct the initial state.  A critical issue
is the initial magnetic field.  Some groups have started with a stellar 
dipole field that threads the disc (Hayashi, Shibata \& Matsumoto 1996, 
Goodson, Winglee \& B\"ohm 1997), 
whilst other groups have taken into account
that the disc may possess a magnetic field on its own (Rast\"atter \& Neukirch
1997, Miller \& Stone 1997).
The simulations by Hayashi et al. \shortcite{hayashi96} and 
Goodson et al. \shortcite{goodson97} both show that
the stellar magnetic field is
wound up by the Keplerian accretion disc and becomes unstable.  A reconnection
event follows and mass is thrown out more or less along the rotational axis.
Goodson et al. suggest that this is the matter forming the stellar jet, but 
they notice that there is also a slower outflow between the jet and the
disc, which may form a disc wind.  In general the simulations show more 
evidence for outflow than for accretion onto the star, though Miller \& Stone
\shortcite{miller:stone} find a polar accretion flow when their disc contains 
a vertical magnetic field, and likewise do Rast\"atter \& Neukirch 
\shortcite{rastatter:neukirch} find that reconnection causes matter to be
accelerated along the field lines leading to the polar caps.

To summarise, the simulations are not yet mature enough to provide any 
information on the structure of steady accretion discs around magnetic stars.
The main reason for this is that they have concentrated on following transient
events for a brief time interval.
However the simulations suggest
that the existence of a dynamo-generated magnetic 
field in the accretion disc is important for the interaction between the disc
and the stellar magnetosphere.  The purpose of this paper is to explore how 
the dynamo may affect the exchange of angular momentum between the accretion
disc and the star.  I compare the torque generated by the winding up of the 
stellar magnetic field with the torque generated by the coupling between a
stellar magnetic field and a dynamo-generated magnetic field in the disc in
Sect. 2.  The existence of two different forms of magnetic torques may influence
the spin evolution of
the X-ray pulsars as suggested in Sect. 3.  Finally my conclusions are 
summarised in Sect. 4.

\section{Magnetic torques}

We can distinguish between two different ways in which the accretion disc and
the star exchanges angular momentum; matter flowing over from the disc to
the star carries with it angular momentum, and the magnetic stress at the
disc surface exerts a torque on the disc.  
The torque due to the angular momentum carried by the accreting matter from the
inner edge of the disc is
\begin{equation}
  \tau_{\rm accr} = \dot M \left(GM r_0\right)^{1/2},
\label{accr_torq}
\end{equation}
where $\dot M$ is the accretion rate, $G$ the gravitational constant, $M$ the
mass of the accreting star, $r_0$ the inner radius of the disc.  
For the rest of this paper it is useful to write 
\begin{equation}
  r_0 = \xi r_{\rm A},
\label{r_0}
\end{equation}
where $r_{\rm A}$ is the Alfv\'en radius
\begin{equation}
  r_{\rm A} = \left(\frac{2\pi^2 \mu^4}{GM \dot M^2 \mu_0^2}
\right)^{1/7},
\label{alfven}
\end{equation}
$\mu_0$ is the permeability of free space, and $\mu$ is the magnetic
dipole moment of the accreting star.  To calculate $\xi$ a detailed model of
the accretion disc is needed \cite{ghosh:lamb2}, but its value is typically
around 0.5 (e.g. Frank, King \& Raine 1992).  The torque can now be written as
\begin{equation}
  \tau_{\rm accr} = \left(2\pi^2\right)^{1/14} \xi^{1/2} \tau_0,
\end{equation}
where
\begin{equation}
  \tau_0 = \left[\frac{\left(GM\dot M^2\right)^3\mu^2}{\mu_0}\right]^{1/7}.
\label{accr_xi}
\end{equation}

The magnetic
torque is the result of the coupling between the vertical magnetic field of
the star and the toroidal magnetic field in the disc.  The torque acting
on the 
upper surface of the disc can be written as
\begin{equation}
  \tau_{\rm mag} = 2 \pi \int_{r_0}^\infty r \frac{B_z B_\phi}{\mu_0} r
\mbox{d}r,
\end{equation}
where $B_z$ is the vertical magnetic field, $B_\phi$ the toroidal field and
$\mu_0$ the magnetic permeability of free space.  There is a similar 
contribution, but with the opposite sign from the lower surface, thus 
angular momentum is exchanged between the disc and the star only if $B_\phi$
changes sign from the upper to the lower surface.  This is true if $B_\phi$
is generated by the winding up of the stellar magnetic field, but it is not true
for a quadrupolar magnetic field generated by a disc dynamo.
Linear mean-field $\alpha\Omega$-dynamos
with a positive $\alpha$-effect (e.g. Stepinski \& Levy 1990, Torkelsson \&
Brandenburg 1994a) generate preferentially quadrupolar magnetic fields, 
but
nonlinear $\alpha\Omega$-dynamos
can generate a large range of different magnetic field configurations
\cite{torkelsson:brandenburg94b}.
It is not necessary that $B_\phi$ is coherent across the entire accretion disc,
because, as we will see later, the magnetic torque is concentrated to region
inside $2 r_0$.

\subsection{The torque due to the winding up of the stellar field in the disc}

\label{winding}

A toroidal field is generated in the disc because
the angular velocities of the disc and the star match only at the co-rotation
radius, $r_{\rm co}$.  Inside $r_{\rm co}$ the disc rotates 
faster than the star so that $B_z B_\phi < 0$ at the upper surface of the disc,
and the disc is losing angular
momentum to the star, whilst $B_z B_\phi > 0$ outside $r_{\rm co}$ and
the disc is gaining angular momentum from the star.  To calculate $B_\phi$
one must know the magnetic diffusivity in the disc (e.g. Campbell \& Heptinstall
1998, Brandenburg \& Campbell 1998).
A simpler approach is to write $B_\phi = \gamma B_z$, where the
azimuthal pitch $\gamma \sim 3$ 
is assumed to be constant (e.g. Ghosh \& Lamb 1979a). 
Obviously this approach must fail close to $r_{\rm co}$, where the 
shear vanishes.
Neglecting this technical complication I calculate the
torque acting on the star due to the interaction between the stellar dipole 
field and the disc
\begin{eqnarray}
  \tau_{\rm mag} = 2\times 2\pi \left\{\int_{r_0}^{r_{\rm co}}\frac{\gamma\mu^2}
{\mu_0 r^6} r^2 \mbox{d}r - \int_{r_{\rm co}}^\infty \frac{\gamma\mu^2}
{\mu_0 r^6} r^2 \mbox{d}r\right\} = \nonumber \\ 
\frac{4 \pi \gamma \mu^2}{3 \mu_0 r_0^3}
\left[1-2\left(\frac{r_0}{r_{\rm co}}\right)^3\right],
\label{mag_torq}
\end{eqnarray}
where $\mu$ is the magnetic dipole moment.  Substituting $r_0$ from 
Eq. (\ref{r_0})
we get
\begin{equation}
  \tau_{\rm mag} = \frac{2\left(16 \pi\right)^{1/7}}{3} \gamma \xi^{-3}
\left(1-2\omega_{\rm s}^2\right) \tau_0,
\label{mag_xi}
\end{equation}
where the fastness parameter $\omega_{\rm s}^2 = (r_0^3/GM)/(r_{\rm co}^3/GM)$
\cite{elsner:lamb} is introduced.

\subsection{The torque due to the coupling between the stellar dipolar magnetic
field and a dynamo-generated toroidal field in the disc}

\label{dynamo}

It is now widely believed that the accretion is driven by a magnetic stress that
is generated by a dynamo in the accretion disc (e.g. Brandenburg et al. 1995,
Stone et al. 1996, Matsumoto \& Tajima 1995). The torque 
needed to drive the accretion is
\begin{equation}
  \dot M \sqrt{GMr}.
\label{acc_torque}
\end{equation}
The magnetic torque, on the other hand, can be written as 
\begin{equation}
  2\pi r 2H r \frac{B_\phi B_r}{\mu_0},
\label{mag_torque}
\end{equation}
where $2H$ is the thickness of the disc.
I write $B_\phi = \gamma_{\rm dyn} B_r$, where $\gamma_{\rm dyn} \sim 10$
\cite{brandenburg}.
Equating Eqs. (\ref{acc_torque}) and (\ref{mag_torque}) the dynamo-generated 
toroidal field is
\begin{equation}
  B_\phi = \sqrt{\frac{\dot M \mu_0 \gamma_{\rm dyn}}
{4 \pi r^2 H}}\left(GMr\right)^{1/4},
\end{equation}
which falls off more slowly with $r$ than the field
that is generated by the winding up of the stellar dipole field in Sect. 
\ref{winding}.
This may be a lower limit to $B_\phi$ as the work by Brandenburg \& Campbell
\shortcite{brandenburg:campbell} suggests that the torque in Eq. 
(\ref{mag_torque}) must be much larger than the torque in Eq. (\ref{acc_torque})
to re-distribute the angular momentum exchanged with the star.

The Shakura-Sunyaev \shortcite{shakura:sunyaev} model predicts that $H/r \propto
r^{1/8}$ for a geometrically thin, optically thick disc in which the gas 
pressure dominates over the radiation pressure.  Because this is sensitive to
the opacity and equation of state,
and to keep things simple, I will take $H/r$ to be a constant.  The torque
due to the coupling between the stellar field and the dynamo-generated toroidal
field in the disc is then
\begin{eqnarray}
  \tau_{\rm mag,dyn} = 2\times 2\pi \int_{r_0}^\infty r \frac{1}{\mu_0}
\frac{\mu}{r^3}B_\phi r \mbox{d} r = \nonumber \\
\frac{4\sqrt{3}}{5} \left(\frac{H}{r}\right)^{-1/2}
\sqrt{\frac{4\pi \gamma_{\rm dyn}\mu^2}{3\mu_0 r_0^3}} 
\left(GM\dot M^2r_0\right)^{1/4}
\label{dyn_torq}
\end{eqnarray}
of either sign.
This can be re-written using Eq. (\ref{r_0}) as
\begin{equation}
 \tau_{\rm mag,dyn} = \frac{4}{5} \left(2^{23} \pi^4\right)^{1/28} 
\gamma_{\rm dyn}^{1/2} \xi^{-5/4} \left(\frac{H}{r}\right)^{-1/2}
\tau_0
\label{dyn_xi}
\end{equation}
This is in a sense an upper limit to the torque caused by the disc dynamo,
as it requires the toroidal magnetic field to be aligned across the entire
disc, but it is potentially the largest component of the
total torque, and must therefore be taken into account.



\section{Discussion:  The spin evolution of disc-accreting X-ray pulsars}

X-ray pulsars are the best laboratories for studying the exchange of angular
momentum between the accretion disc and the accreting star, because the moment
of inertia of a neutron star is comparatively small, and an X-ray pulsar can
be timed accurately.  The torque is measured from the change in spin frequency
\begin{equation}
 \dot \nu = \frac{\tau}{2\pi I},
\end{equation}
where $I$ is the moment of inertia of the neutron star.
Adding Eqs. (\ref{accr_xi}), (\ref{mag_xi}) and (\ref{dyn_xi}) 
the spin change is
\begin{equation}
  \dot \nu = 2.4\times 10^{-14} \,\,\mbox{Hz\,\,s}^{-1} I_{40}^{-1} m^{3/7}
\dot M_{14}^{6/7} \mu_{20}^{2/7} \tilde \tau,
\label{total}
\end{equation}
where
\begin{equation}
  \tilde \tau =
1.2 \xi^{1/2} + 1.2 \frac{\gamma}{\xi^3} \left(1- 2\omega_{\rm s}^2\right)
\pm 1.7 \left(\frac{\gamma_{\rm dyn}}{\xi^{5/2}}\right)^{1/2}\left(\frac{H}{r}
\right)^{-1/2},
\end{equation}
$I_{40}= I/10^{40}$ kg\,m$^2$, $m = M/M_{\sun}$, $\dot M_{14} = \dot M/10^{14}$ 
kg\,s$^{-1}$, and $\mu_{20} = \mu /10^{20}$ T\,m$^3$. 
The Ghosh \& Lamb \shortcite{ghosh:lamb3} model, which corresponds to the first
two terms above, 
predicts that 
at a sufficiently low accretion rate the torque is spinning down the neutron
star, but as the accretion rate increases and the inner disc edge is pushed
closer to the neutron star, the torque changes sign, and the
spin up torque increases with increasing accretion rate.  At high accretion
rates
\begin{equation}
  \dot \nu \propto \frac{\dot M^{6/7}}{\nu^2}.
\end{equation}
It is
difficult to derive any form of similar relation when the angular momentum 
exchange is dominated by the coupling to a dynamo-generated magnetic field in
the disc.

The amount of timing data of X-ray pulsars has increased dramatically 
since the launch of BATSE/CGRO.  Bildsten et al. \shortcite{bildsten97} have
recently presented a compilation of 5 years of monitoring of X-ray pulsars.
Most X-ray pulsars are fed by stellar winds from their supergiant companions,
and are therefore irrelevant for this paper.
There are two groups of X-ray pulsars that have accretion discs.  These
are the Be/X-ray transients and an inhomogeneous group of binaries with steady
accretion discs.  The neutron stars in the Be/X-ray transients sometimes
pick up an accretion disc at the time of their periastron passage, which leads
to an outburst of X-rays.
The observations of transient X-ray
pulsars, such as EXO 2030+375 (e.g. Reig \& Coe 1998) and A0535+262 
give some support to the Ghosh \& Lamb model as the X-ray pulsars spin up
during the outbursts, and that the spin-up rate decreases as the X-ray flux
goes down, but a
major uncertainty is the relation between the observed X-ray flux and the
accretion rate \cite{bildsten97}.  

The persistent X-ray pulsars with accretion
discs are comparatively few and make up an inhomogeneous set.  The systems I
will discuss in the following are 4U 1626-67, GX 1+4, Cen X-3 and OAO 1657-415,
in which the mass losing stars are a degenerate dwarf, a red giant, an O6-8
supergiant, and an OB supergiant respectively.  4U 1626-67 had been spinning
up steadily at a rate $\dot \nu = 8.5\times 10^{-13}$ Hz\,s$^{-1}$
from its discovery by {\em Uhuru} \cite{giacconi72} until 
the beginning of the BATSE observations, but it is now spinning
down just as steadily at a rate $\dot \nu = - 7.2\times 10^{-13}$ Hz\,s$^{-1}$
\cite{chakrabarty97}.  GX 1+4 is similar in the sense that it was also 
discovered in a state of spinning up at $\dot \nu = 6.0\times
10^{-12}$ Hz\,s$^{-1}$
(Davidsen, Malina \& Bowyer 1977, Nagase et al. 1989), but since the days of
{\em Ginga} \cite{makishima88} it has been spinning down at $\dot \nu = 
-3.7\times 10^{-12}$ Hz\,s$^{-1}$.
Chakrabarty et al. \shortcite{chakrabarty97b} have studied the correlation
between the pulsed X-ray flux in the 20 - 60 keV band with
the spin-down rate of GX 1+4.  There is no clear
correlation, and the largest spin-down rates seem to occur at
the highest X-ray fluxes, which contradicts the Ghosh \& Lamb model.

Cen X-3 shows an altogether different behaviour.
On average it is spinning up at a rate $\dot \nu = 8\times
10^{-13}$ Hz\,s$^{-1}$,
but BATSE revealed a lot of fine structure in its spin evolution, and it is
alternating between spin-up and spin-down phases with $\dot \nu = 7\times
10^{-12}$
Hz\,s$^{-1}$ and $\dot \nu = -3\times 10^{-12}$ Hz\,s$^{-1}$, respectively 
\cite{bildsten97}.  
Typically it spends 10 to 100 days in one
state before switching to the other state faster than can be resolved by BATSE. 
OAO 1657-415 appears to
be a sister system of Cen X-3's although the companion has not been identified.

The existence of two states of comparable but oppositely directed torques is
not expected from the Ghosh \& Lamb model.  It is, 
though, a natural consequence,
if $\tilde \tau$ is dominated by its last term, the coupling between the stellar
magnetic field and the dynamo-generated field in the disc.  
In that case we expect the
spin-up torque to be somewhat larger than the spin-down torque in agreement 
with the observations.


There are two time scales that must be explained, the time scale over which
the torque is constant, and the time scale for reversing the torque.  These
time scales are 10 - 100 days, and less than 10 days, respectively, for Cen X-3.
In the cases of GX 1+4 and 4U 1626-67, on the other hand, the torque remains 
constant for several years, and the sparse sampling at the time of torque 
reversal provides a generous upper limit for the time scale of the torque
reversal.
The time scale for reversing the torque should be the same as the
time scale for reversing the magnetic field via diffusion in my model.  I assume
the diffusive time scale to be comparable to the viscous time scale
\begin{equation}
  t_{\rm vis} \sim \frac{r^2}{\nu_{\rm turb}} = \alpha_{\rm SS}^{-1} 
\left(\frac{H}{r}\right)^{-2} \Omega^{-1},
\end{equation}
where $\nu_{\rm turb} = \alpha_{\rm SS} H c_{\rm s}$ is the turbulent
viscosity, $\alpha_{\rm SS}$
the Shakura-Sunyaev \shortcite{shakura:sunyaev} parameter, and $c_{\rm s}$ the
sound speed.  The dynamo torque is so strongly concentrated to the inner part
of the disc that half of the torque is due to the disc inside
$1.74 r_0$.  The viscous time scale at this radius is typically less than
0.5 days.  

It is much less clear what sets the time scale over which the
magnetic torque is constant. 
The ratios of the intervals between torque reversals and
the orbital periods may be roughly comparable for Cen X-3 and GX 1+4, though
only lower limits to the orbital period of the latter are known 
\cite{chakrabarty:roche}.  The time interval between the torque reversals 
is comparable to the viscous time scale for the entire disc, which
suggests that the time scale is set by the time over which the disc can
support a given field configuration against diffusion.  This speculation 
fails completely in the case of 4U 1626-67, which is the smallest of
the systems I have discussed, and yet the torque remains constant over 
a time scale of several years.

Some other mechanisms have been proposed to explain the transitions between
spin-up and spin-down states in X-ray pulsars.  van Kerkwijk et al. 
\shortcite{vankerkwijk98} 
note that the accretion disc can be subject of a warping
instability due to the irradiation from the neutron star.  This warp may be
so extreme that the inner part of the accretion disc flips over and rotates in
the opposite direction, which would lead to a torque reversal.  Yi, Wheeler \&
Vishniac 
\shortcite{yi97} suggest a modification of the Ghosh \& Lamb model.  The torque
reversals are due to small changes in the accretion rate, but
the inner part of the accretion disc changes between a standard thin
disc and an advection-dominated flow at the same time.  
The advection-dominated flow is less
efficient in radiating energy, which can solve the problem of the lack of
a correlation between the torque and the observed X-ray flux.  A third 
possibility was suggested by Nelson et al. \shortcite{nelson97}.  The spin-down
can be explained if the disc is retrograde, which was first suggested by
Makishima et al. 
\shortcite{makishima88}.  It is difficult to see how a retrograde
disc can appear in a Roche-lobe overflow, but numerical simulations of accretion
from winds have shown that a temporary disc rotating in the 'wrong' direction
may appear (e.g. Fryxell \& Taam 1988).

\section{Summary}

I have shown that the torque acting between an accretion disc and an accreting
star can be enhanced by the presence of an intrinsic magnetic field in the
accretion disc.
The orientation of the magnetic field in the disc, and thus the direction of
the magnetic torque between the disc and the star, is
arbitrary, because the dynamo does not have any information on the rotation of
the neutron star.
In
particular it is therefore possible to reverse the torque by reversing the
magnetic field in the disc.  This mechanism may explain the observed torque
reversals in some X-ray pulsars that are fed by accretion discs.  The short
time scale for the reversals compared to the long time scale over which the
torque remains the same may be explained as the difference between the diffusive
(viscous) time scales for the inner region of the disc, to which the torque
is concentrated, and the entire accretion disc.  A problem with this 
connection is that it cannot explain the stability of the torque in the
short-period X-ray binary 4U 1626-67.

\section*{Acknowledgements}

I acknowledge the support of an EU post-doctoral fellowship, and I thank 
Axel Brandenburg for reading a draft version of this paper.

\end{document}